\documentclass[review]{elsarticle}

\usepackage{lineno,hyperref}
\usepackage[margin=25mm]{geometry}
\usepackage{siunitx}
\usepackage{amsmath,amssymb}
\usepackage{graphicx,xcolor}
\usepackage{lineno}
\modulolinenumbers[2]

\newcommand{\nNu}{\mathord{\mathit{Nu}}}
\newcommand{\nRe}{\mathord{\mathit{Re}}}
\newcommand{\nPr}{\mathord{\mathit{Pr}}}
\newcommand{\degC}{\si{\degreeCelsius}}

\journal{International Journal of Multiphase Flow}

\biboptions{authoryear}

\begin{document}

\begin{frontmatter}

\title{Experimental study of cavitating flow influenced by heat transfer 
from heated hydrofoil}

\author[IFS]{Junnosuke Okajima\corref{mycorrespondingauthor}}
\cortext[mycorrespondingauthor]{Corresponding author}
\ead{j.okajima@tohoku.ac.jp}

\author[IFS,ENG]{Masaki Ito}
\author[IFS]{Yuka Iga}

\address[IFS]{Institute of Fluid Science, Tohoku University, 2-1-1 Katahira, 
Aoba-ku, Sendai, Miyagi 980-8577, Japan}
\address[ENG]{Mechanical Engineering Division, School of Engineering, 
Tohoku University, 6-6, Aramaki Aza Aoba, Aoba-ku, Sendai, Miyagi 980-8579, 
Japan}

\begin{abstract}
This study experimentally investigated the influence of heat transfer from a heated hydrofoil on cavitating flow to understand the evaporation phenomenon under high-heat-flux and high-speed conditions. 
A temperature difference was generated between the hydrofoil and mainstream by installing an aluminum nitride heater in a NACA0015 hydrofoil fabricated from copper.
A cavitation experiment was performed in a high-temperature water cavitation tunnel at the Institute of Fluid Science, Tohoku University. 
The effect of heating on cavitating flow was evaluated by changing the mainstream velocity and pressure, namely the cavitation number, at a fixed heater power of 860 W.
Results showed that the heat transfer from the hydrofoil affected cavitating flow in terms of the cavity length, cavity aspect, and periodicity.
The effect on the cavity length became stronger at a lower velocity owing to a higher hydrofoil temperature.
The variation in periodicity implied that the heating effect reduced the unsteadiness of cavitation.
A modified cavitation number was proposed by considering the heat transfer from the heated wall.
A thermal correction term was derived by considering that the fluid temperature close to the heated hydrofoil was affected by the turbulent convective heat transfer between the mainstream and hydrofoil.
The corrected cavitation number can be considered as a parameter that describes the cavity length in the isothermal and heated cases in a unified expression.
\end{abstract}


\end{frontmatter}

\section{Introduction}
In recent years, the improvement of the cooling ability of high-power-density electrical devices and systems~\citep{Mudawar2011, Nishimura2020} has become an important problem.
Cooling ability is also critical to nuclear plants~\citep{Wang2021}, rocket engines~\citep{Mudawar2011}, and surgical devices in the medical field~\citep{Okajima2014,Okajima2019}.
Numerous researchers have proposed methods such as modification of heat transfer surfaces~\citep{Sajjad2021}, utilization of flash boiling to reduce temperature~\citep{Okajima2014,Okajima2019}, and boiling in high-speed flow fields\citep{Trejo2016, Ning2019} to improve cooling ability.
Although high-speed flow increases the cooling ability, it leads to the occurrence of cavitation in the flow passage, which culminates in harmful effects such as vibration, noise, and erosion, as reported by \cite{Ning2019}.
However, the interaction between phase change phenomena and high-speed flow has not been investigated sufficiently in the field of boiling flow.
Therefore, to further improve the cooling system while simultaneously achieving better performance and undamaged operation, we have to understand the relationship between the unsteady-dynamic behavior of a flow field and phase change phenomena under intense heat flux from a heated wall. For this reason, in this study, we focused on cavitating flow with a heated wall.

Before discussing the cavitating flow with a heated wall, the thermodynamic self-suppression effect, which is a known thermal phenomenon in cavitation, is introduced.
The thermodynamic self-suppression effect~\citep{Stepanoff1964} causes local temperature variation in cryogenic fluids~\citep{Yoshida2006, Le2019}, Freon~\citep{Franc2004}, and high-temperature water~\citep{Kato1996}.
The mechanism of the thermal self-suppression effect can be explained as follows: the sensible heat in the liquid around the vapor bubble is consumed by evaporation, and the temperature of the liquid-vapor interface decreases; subsequently, the saturation pressure decreases, suppressing the evaporation. 
This effect is preferable for the fluid machinery because the cavity is suppressed.
In this phenomenon, the heat exchange between a fluid and a solid wall is insignificant because the magnitude of temperature variation is sufficiently small.
Therefore, this phenomenon occurs because of heat transfer inside the fluid.

Meanwhile, heat transfer from a heated wall also affects cavitation.
Under this heating process, the phenomenon should be close to the boiling flow.
The cavitating flow with a heated object can be considered similar phenomena to the subcooled boiling in high-speed flow.
Several studies have examined cavitation around heated objects.
\cite{Arakeri1973} used a heated object to visualize a viscous boundary layer. 
They increased the object temperature by 2 K compared to the mainstream temperature. 
They confirmed that this temperature rise did not affect the characteristics of the viscous boundary layer.
\cite{Wang2015} evaluated the influence of the temperature of an object on the cavitating flow and drag.
They considered the temperature dependency of saturation pressure and latent heat in a cavitation model for numerical simulation and used conjugated heat transfer boundary conditions at a solid/liquid interface.
However, they did not discuss the heat transfer process of cavitating flow close to a wall.
\cite{Kimoto1989} evaluated the Nusselt number around a heated circular cylinder with the cavitating flow.
They reported that the Nusselt number increased as the cavitation number decreased. However, when supercavitation occurred, heat transfer degraded compared to single-phase flow.
From the viewpoint of heat transfer, \cite{Lorenzini2019} developed a microfluidic cooling device and found that vapor generation was promoted at the rear part of pin fins owing to the local low pressure caused by flow separation.
\cite{Schneider2006} evaluated the enhancement of heat transfer by cavitating flow in a microchannel and discussed the relationship between the heat transfer mechanism and flow pattern.
These studies primarily focused on heat transfer performance. However, the dynamic variation in the characteristics of cavitating flow caused by heating has not been investigated in detail.

As mentioned above, to simultaneously achieve superior performance and undamaged operation in a cooling system, the relationship between the unsteady-dynamic behavior of a flow field and phase change phenomena under intense heat flux from a heated wall should be clarified.
Consequently, a unified understanding of the cavitation in a heated object and subcooled boiling in high-speed flow is expected.
We approach this phenomenon from the cavitation viewpoint.

The objective of this study is to understand the evaporation phenomenon in terms of high-speed flow and high heat flux to evaluate the influence of the heat transfer from a heated object on cavitating flow.
We used a NACA0015 hydrofoil, which is generally used in cavitation experiments. The cavity length was selected as an evaluation parameter, and the cavitation number was corrected by considering the heat transfer effect.

\section{Experimental system}
\subsection{Cavitation tunnel}
\begin{figure} [htbp]
\centering
	\includegraphics[width=3.34in]{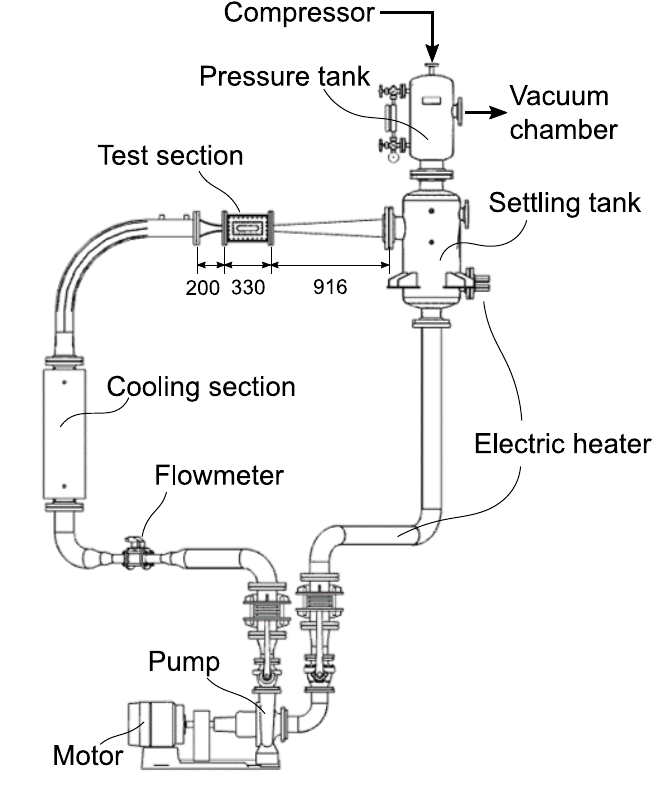}
\caption{Schematic of high-temperature water cavitation tunnel}
\label{fig_tunnel}
\end{figure}

The high-temperature water cavitation tunnel, which is a closed-type tunnel, as shown in Fig. \ref{fig_tunnel}, at the Institute of Fluid Science, Tohoku University, was used for this study.
The details of the cavitation tunnel have been presented in \cite{Iga2016}, \cite{Kang2019}, and \cite{Hanyuda2021}.

The test section has a rectangular-shaped cross-section with a height of 30 mm, width of 20 mm, and length of 330 mm.
The nozzle and diffuser were installed on the upstream and downstream sides, respectively.
The nozzle contracts the flow channel from 114 mm $\times$ 114 mm to 20 mm $\times$ 30 mm in the length of 197 mm.
The diffuser enlarges the flow channel gradually to 108 mm $\times$ 108 mm in the length of 855 mm.
The centrifugal pump (100X80IFWM, Ebara Co. Ltd.) circulates water as a working fluid up to 700 L/min.
Two electrical heaters, 10 kW in total, were installed, and the mainstream temperature was measured by a platinum resistance thermometer 100 $\mathrm{\Omega}$ (Nihondensoku, Co. Ltd.), which was installed at the exit of the diffuser, with an accuracy of JIS CLASS B.
The volumetric flow rate was measured by an electromagnetic flowmeter (LF410, Ryutaikogyo, Co. Ltd.) with an accuracy of $\pm$ 0.5 \%
The pressure inside the tunnel varied between 0.02 MPa and 0.6 MPa, and the mainstream temperature was controlled between room temperature and 140 \degC. 
The flow rate and pressure can be controlled independently; therefore, the experiment under various cavitation numbers can be operated.

\begin{figure}[htbp] 
\centering
	\includegraphics[width=5.34in]{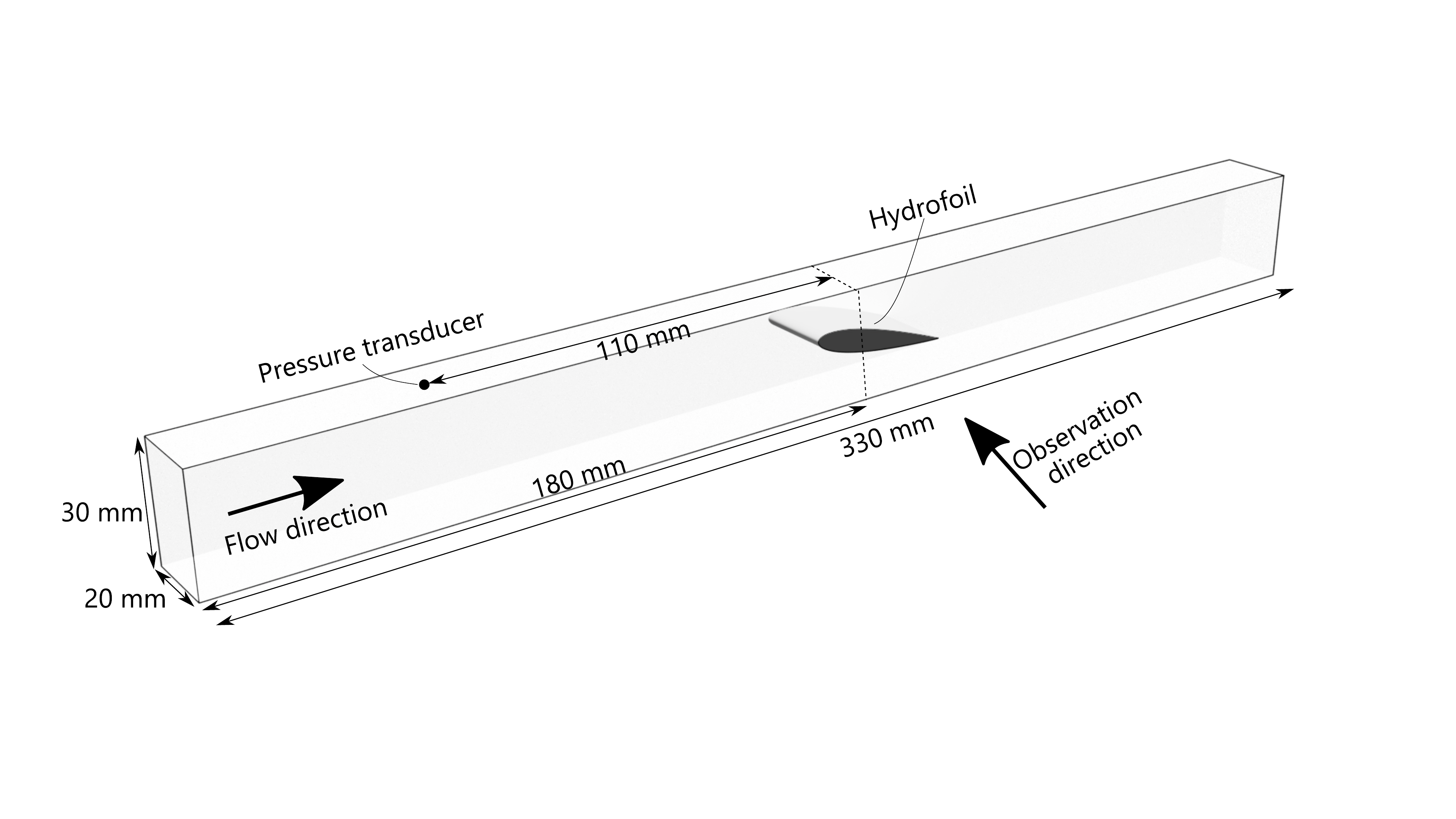}
\caption{Schematic of test section}
\label{fig_testsect}
\end{figure}

Figure \ref{fig_testsect} shows the schematic of the test section.  
The mainstream pressure was measured using a pressure transducer (PH-10KB, Kyowa Co. Ltd.), which was installed 110 mm upstream from the center of the test section.
The pressure transducer was connected to the strain amplifier (DPM-911B, Kyowa Co. Ltd.), and the total accuracy of pressure measurement is $\pm$ 0.5 \%.
The cavity aspect was recorded using a high-speed camera (FASTCAM Mini AX50, Photron Co. Ltd.) 
with the two metal halide lamps (HVC-UL, Photron Ltd.) that illuminated the hydrofoil from the upstream and downstream sides to reduce the shadow.
The frame rate and exposure time were 2000 fps and 10 $\si{\us}$, respectively, 
and 2,000 images were acquired by recording for 1 s.

\subsection{Hydrofoil}
\label{sect_hydrofoil}
\begin{figure}[htbp] 
\centering
	\includegraphics[width=4.34in]{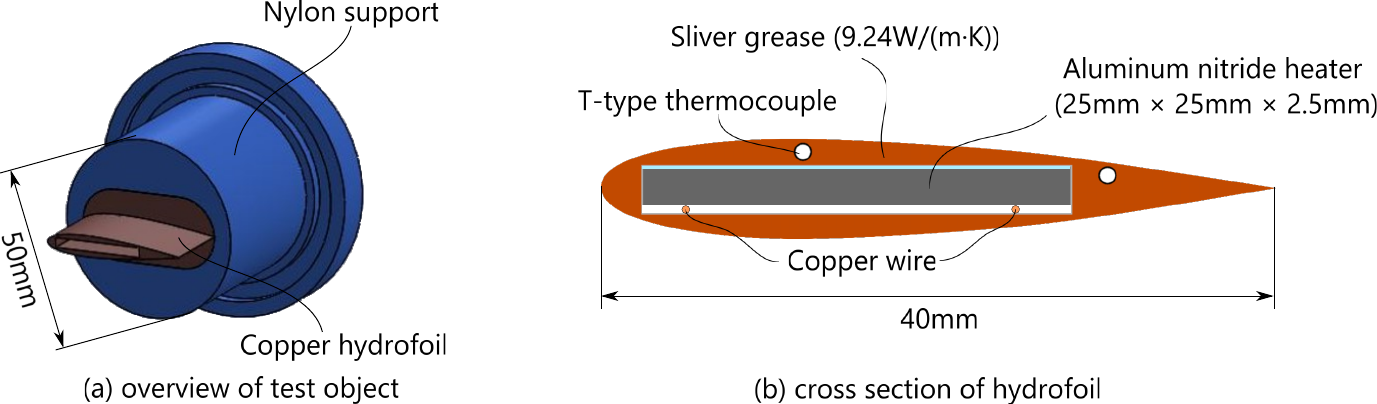}
\caption{Schematic of hydrofoil with heater}
	\label{fig_hydrofoil}
\end{figure}

As shown in Fig. \ref{fig_hydrofoil}, the NACA0015 hydrofoil fabricated from pure copper was used. The chord length and span of the hydrofoil were 40 mm and 20 mm, respectively.
The hydrofoil was installed in the test section using a supporter fabricated from nylon, which reduced the heat loss from the hydrofoil to the tunnel wall.
The test object shown in Fig. \ref{fig_hydrofoil}(a) was fixed in the tunnel wall.
A 25 mm $\times$ 25 mm $\times$ 2.5 mm aluminum nitride heater (WALN-1, Sakaguchi E.H Voc Corp.), with a maximum power of 900 W as the catalog value, was inserted in the hydrofoil through a rectangular hole.
Silver grease was applied at the contact interface to reduce thermal contact resistance.
The electrical power supplying the heater was controlled using a volt slider (S-260-5, Yamabishi Electric Co., Ltd.) connected to 200 VAC.
Additionally, a T-type thermocouple was inserted into the hydrofoil immediately below the suction surface.
The actual surface temperature of the hydrofoil is appropriate for evaluating the heating effect. However, it is difficult to measure this temperature and its distribution on a copper surface.
Therefore, in this study, the value measured by the thermocouple was defined as the representative temperature of the hydrofoil, $T_{w}$.
In addition, the temperature difference, $\Delta T_{w}$, was defined as the difference between the hydrofoil temperature and mainstream temperature, $T_{\infty}$.

\subsection{Experimental conditions}
In the experiment, the angle of attack was fixed as \ang{12}.
Considering the mainstream temperature, the thermodynamic self-suppression effect becomes remarkable in high-temperature water.
Our previous study \citep{Hanyuda2021} shows that the suppression effect appeared in the water at more than 100 \degC.
Therefore, 90 \degC was selected as the mainstream temperature in this study. 
The heater was operated at 200 VAC, and a heat generation of 860 W was achieved in this experiment.
Hence, the heater power was constant at 860 W.
The cavitation number is defined by the following equation:

\begin{equation}
	\sigma = \frac{p_\infty-p_{sat} \left( T_\infty \right)}{\displaystyle \frac{1}{2}\rho_L U_\infty^2}.
	\label{eq_sigma}
\end{equation}

\noindent
Table \ref{tab_exp_cond} shows the experimental conditions of the inlet pressure and mainstream velocity.
These conditions were selected such that the cavitation number varied from 3.0 to 4.1.
Because the blockage ratio in this experiment is relatively large (33\%), this experiment will not be compared to other experiments on hydrofoil properties with an external flow. However, the difference between the heating and non-heating effects under the same blockage conditions will be evaluated.

In this study, for attached sheet cavitation, cavity length was defined as the distance from the leading edge to the rear edge of the cavity where the stationary cavity exists.
For unsteady cavitation, cavity length was defined as the length of the sheet cavitation where the rear edge of the sheet cavity was clearly closed just before cloud shedding. Under this definition, 10 events were randomly selected from 2000 images, and their cavity lengths were measured.
The period was estimated by counting the number of images to evaluate the cycle. 
The period of unsteady cavitation was defined as the time from the complete disappearance of cavitation to its disappearance again, and 5 events were randomly selected from 2000 images and statistically processed.

\begin{table}[htbp]
\caption{Experimental conditions}
\begin{center}
\label{tab_exp_cond}
\begin{tabular}{|c|c|c|c|}
\hline
 & Condition I & Condition II & Condition III\\
\hline
	mainstream velocity $U_\infty$ [m/s] & 4.5--5.3 & 5.5 (const.) & 6.8--8.0\\
\hline
	mainstream pressure $p_\infty$ [MPa] & 0.110 (const.) & 0.114--0.130 & 0.160 (const.)\\
\hline
	cavitation number $\sigma$ [-] & \multicolumn{3}{c|}{3.0--4.1} \\
\hline
	Reynolds number $\nRe$ [-] & 5.5--6.5$\times 10^5$ & 6.7$\times 10^5$ & 8.4--9.8$\times 10^5$\\
\hline
\end{tabular}
\end{center}
\end{table}

\section{Experimental results}
\subsection{Hydrofoil temperature}
\begin{figure}[htbp] 
\centering
	\includegraphics[width=3.34in]{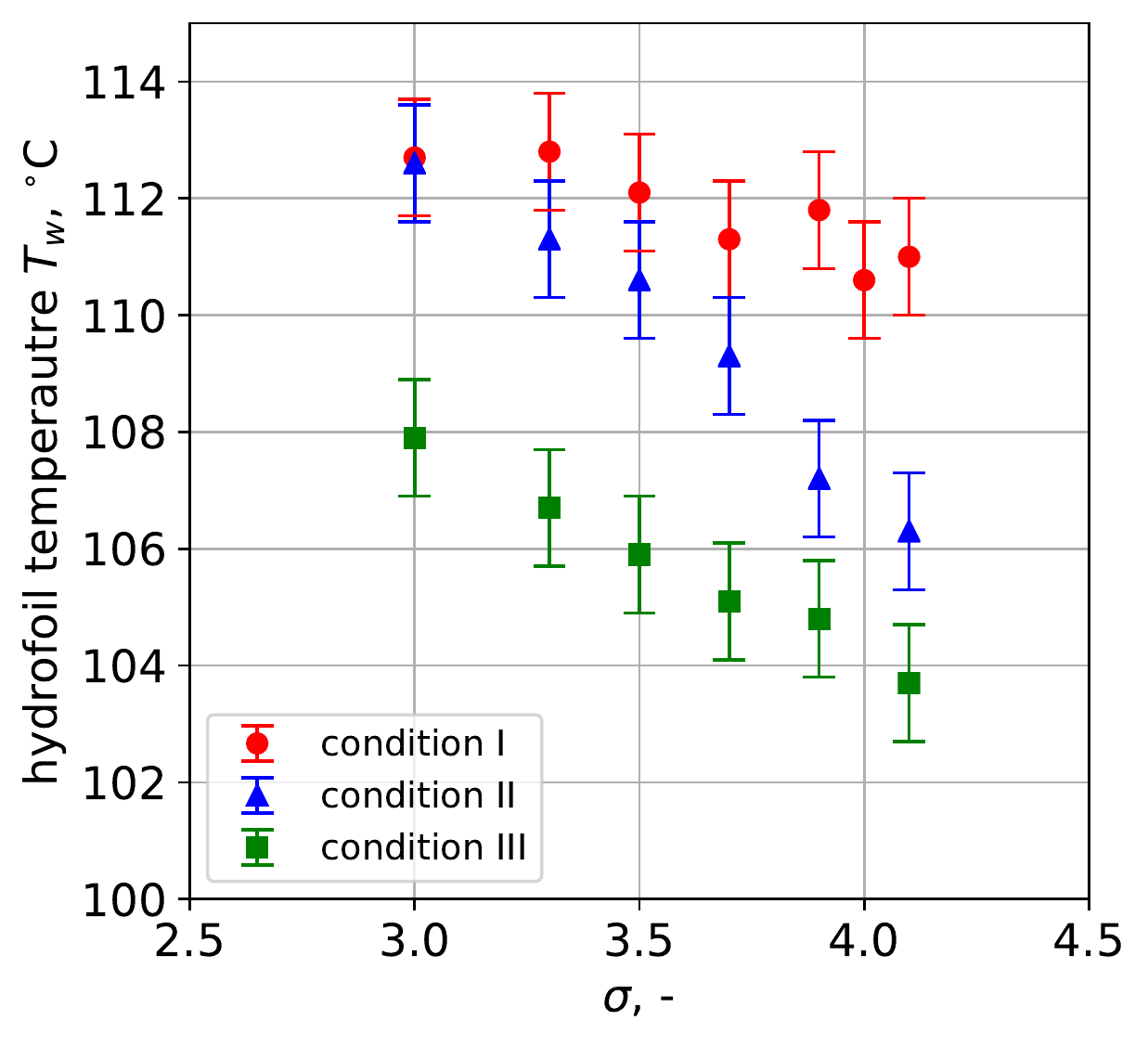}
\caption{Variation in hydrofoil temperature with cavitation number}
\label{fig_sig-T}
\end{figure}

Figure \ref{fig_sig-T} shows the variation in the hydrofoil temperature with the cavitation number. 
As explained in Section \ref{sect_hydrofoil}, the temperature measured by the installed thermocouple was the local temperature at the measurement point.
However, it was considered the representative temperature in this study.
The hydrofoil temperature was determined by the balance between the heater power and convective heat transfer on the pressure and suction surfaces.
Furthermore, the cavitation on the suction surface affected the convective heat transfer.

The hydrofoil temperature decreased as the mainstream velocity increased because the heat transfer from the hydrofoil to the flow was enhanced.
Thus, the hydrofoil temperature in Condition III was the lowest.
Furthermore, in each condition, the hydrofoil temperature increased as the cavitation number decreased because the heat transfer between the hydrofoil and mainstream flow was prevented by cavity development.

\subsection{Observation of cavity aspect}
\begin{figure}[htbp] 
\centering
	\includegraphics[width=6.0in]{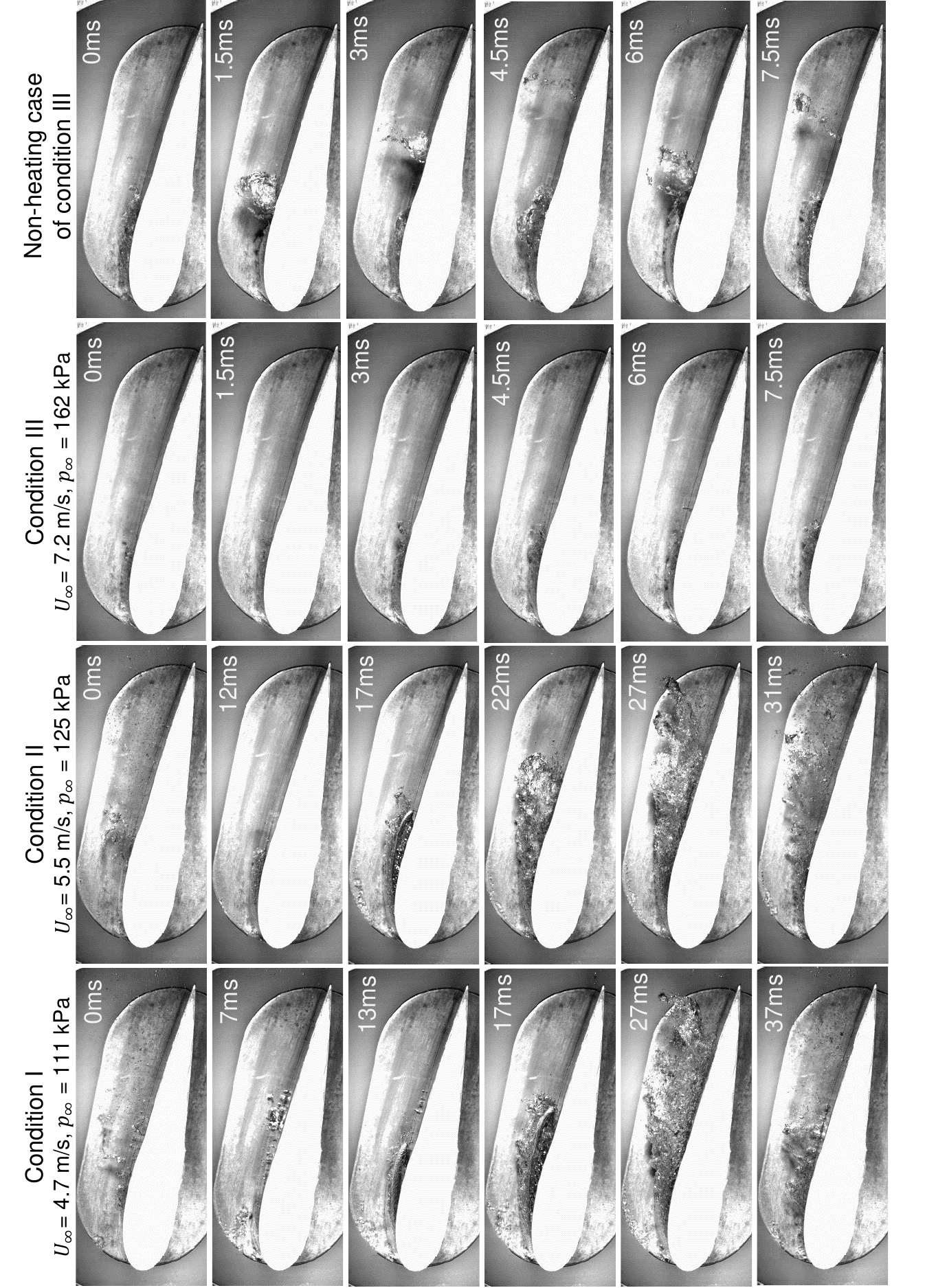}
\caption{Comparison of cavity aspects in each condition at $\sigma$ = 3.7}
\label{fig_flow_pattern}
\end{figure}

Figure \ref{fig_flow_pattern} shows the comparison of the cavity aspect for each experimental condition. 
The flow patterns for $\sigma$ = 3.7 are shown as an example.
Each column in Fig. \ref{fig_flow_pattern} represents one cycle of the cavitation pattern.
The attached sheet cavitation shown in the third column of Fig. \ref{fig_flow_pattern} (Condition III) and sheet/cloud cavitation in the fourth column of Fig. \ref{fig_flow_pattern} (non-heating case of Condition III) were observed in the non-heating and heating cases of Condition III.
In addition, cavitation patterns changed drastically when the heat was supplied in conditions I and II.
At first, a sheet cavity was developed with a clear interface.
After the sheet cavity reached its maximum length, a small bubble cluster separated from the trailing edge of the cavity.
Finally, the cavity disappeared.
In this study, this flow pattern is called the "unsteady cavitation observed in heating conditions".

In particular, in Condition I, bubble generation was observed on the suction side during cavity development, as shown in the images for 7 ms and 37 ms.
These bubbles appeared slightly downstream from the maximum thickness of the suction surface.
The surface temperature in this area appeared to be relatively higher than that in other areas owing to the degradation of heat transfer caused by flow separation.
Vapor bubbles were generated in the high-temperature area on the hydrofoil.

\begin{figure}[htbp] 
\centering
\includegraphics[width=5.0in]{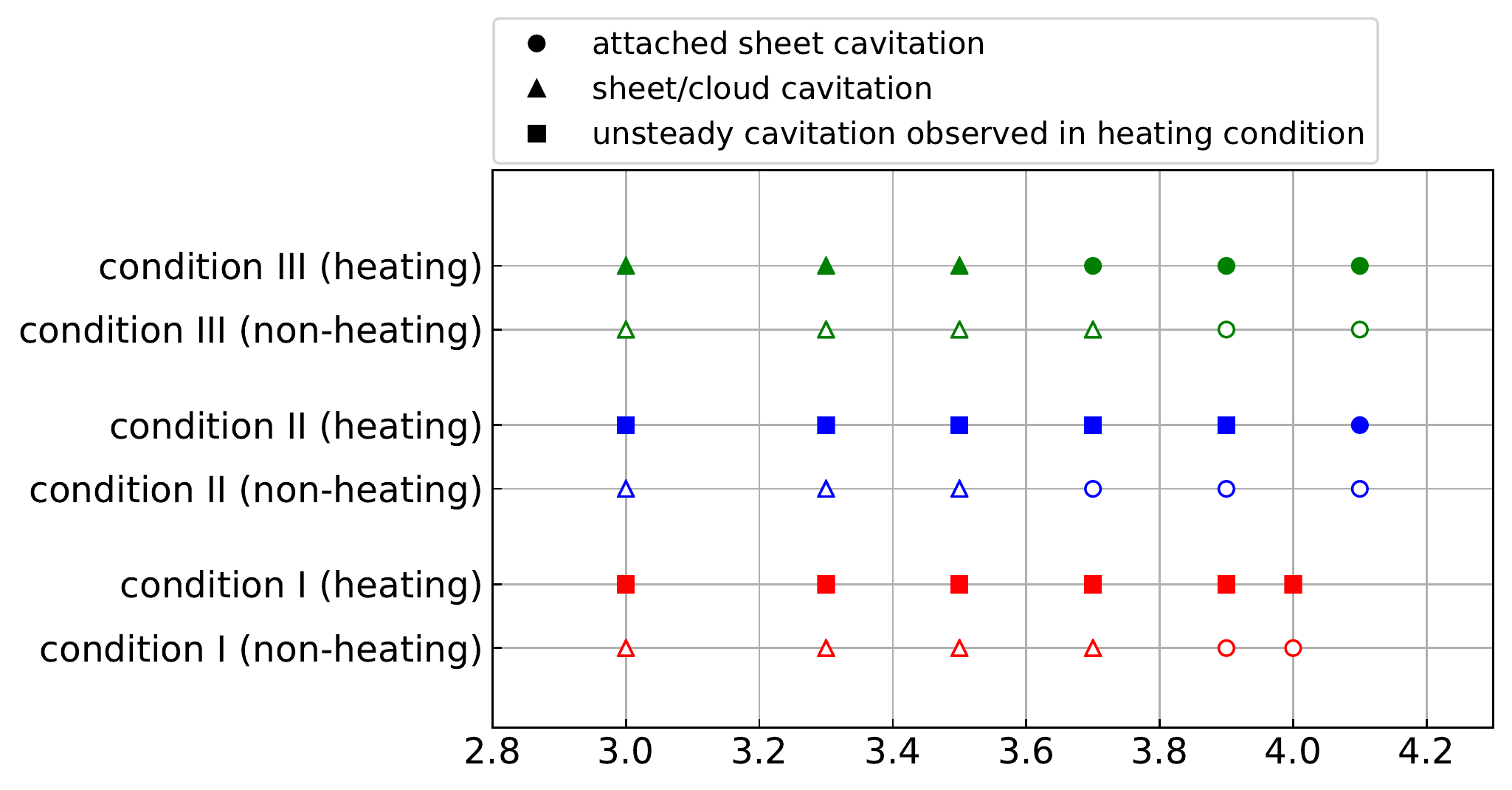}
\caption{Flow pattern map}
\label{fig_cavi_map}
\end{figure}

The observed flow patterns in this study are summarized in Fig. \ref{fig_cavi_map}.
As shown in Fig. \ref{fig_cavi_map}, there is no significant difference among the non-heating cases.
Conversely, in the heating cases, the unsteady cavitation observed in the heating conditions primarily appeared in conditions I and II.
In the heating case of Condition III, typical sheet/cloud cavitation was observed even in heating condition.

\subsection{Cavity length}
\begin{figure}[htbp] 
\centering
	\includegraphics[width=3.34in]{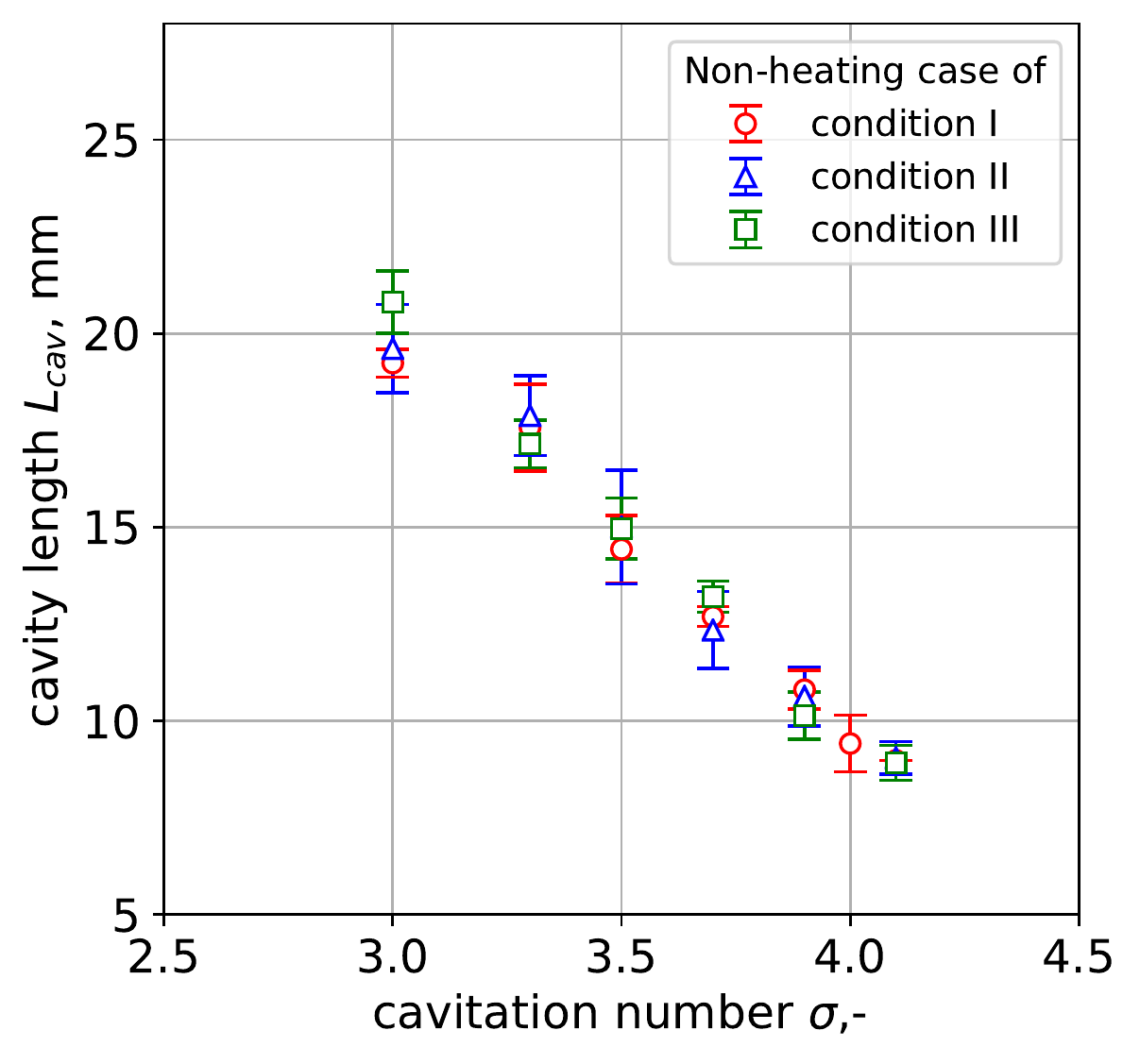}
\caption{Variation in cavity length with cavitation number without 
hydrofoil heating}
\label{fig_L_noheat}
\end{figure}

The cavity length was measured to quantitatively evaluate the cavitation pattern.
Figure \ref{fig_L_noheat} shows the variation in the cavity length with the cavitation number without hydrofoil heating.
The cavity lengths for each condition were similar.
This is a reasonable trend, and a small difference might be caused by the difference in the Reynolds number and mainly the reading error during the measurement of the cavity length.

\begin{figure}[htbp] 
\centering
	\includegraphics[width=3.34in]{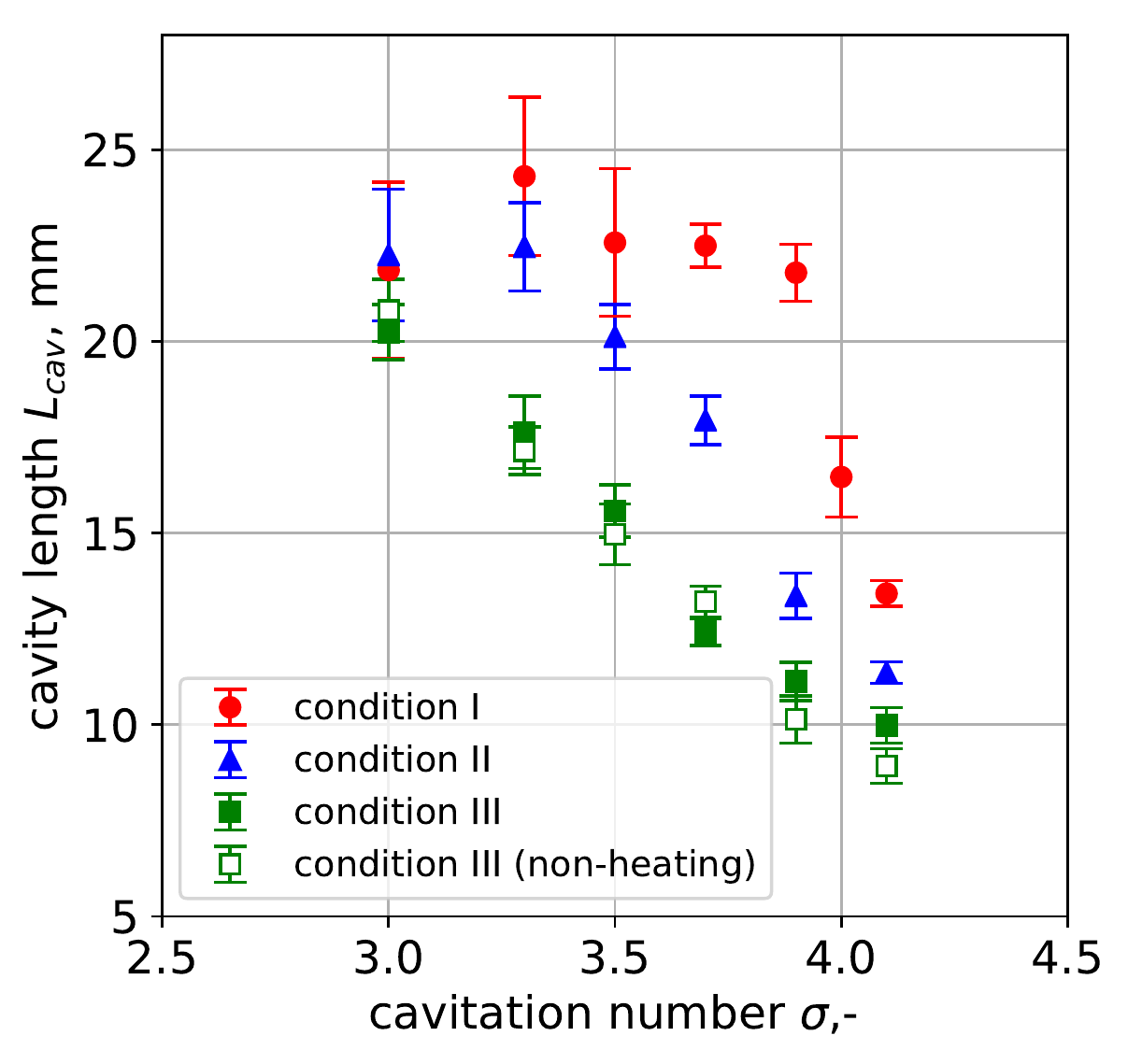}
\caption{Variation in cavity length with cavitation number under heating and non-heating conditions. Solid and hollow symbols represent the heating and non-heating conditions, respectively}
\label{fig_L_heat}
\end{figure}

Figure \ref{fig_L_heat} shows the variation in the cavity length with the cavitation number under heating and non-heating conditions. Different trends were observed for conditions I and II for the non-heating case. 
In particular, the cavity length rapidly increased at approximately $\sigma = 4.0$ and then changed gradually as the cavitation number decreased.
There was a slight difference between Condition III and the non-heating case.
Overall, cavity development differed according to the flow conditions, even though the same heater power was applied.
The cavitating flow in Condition I was the most affected by heating, whereas that in Condition III was not affected by heating.

\subsection{Periodicity in phenomena}

\begin{figure}[htbp] 
\centering
	\includegraphics[width=3.34in]{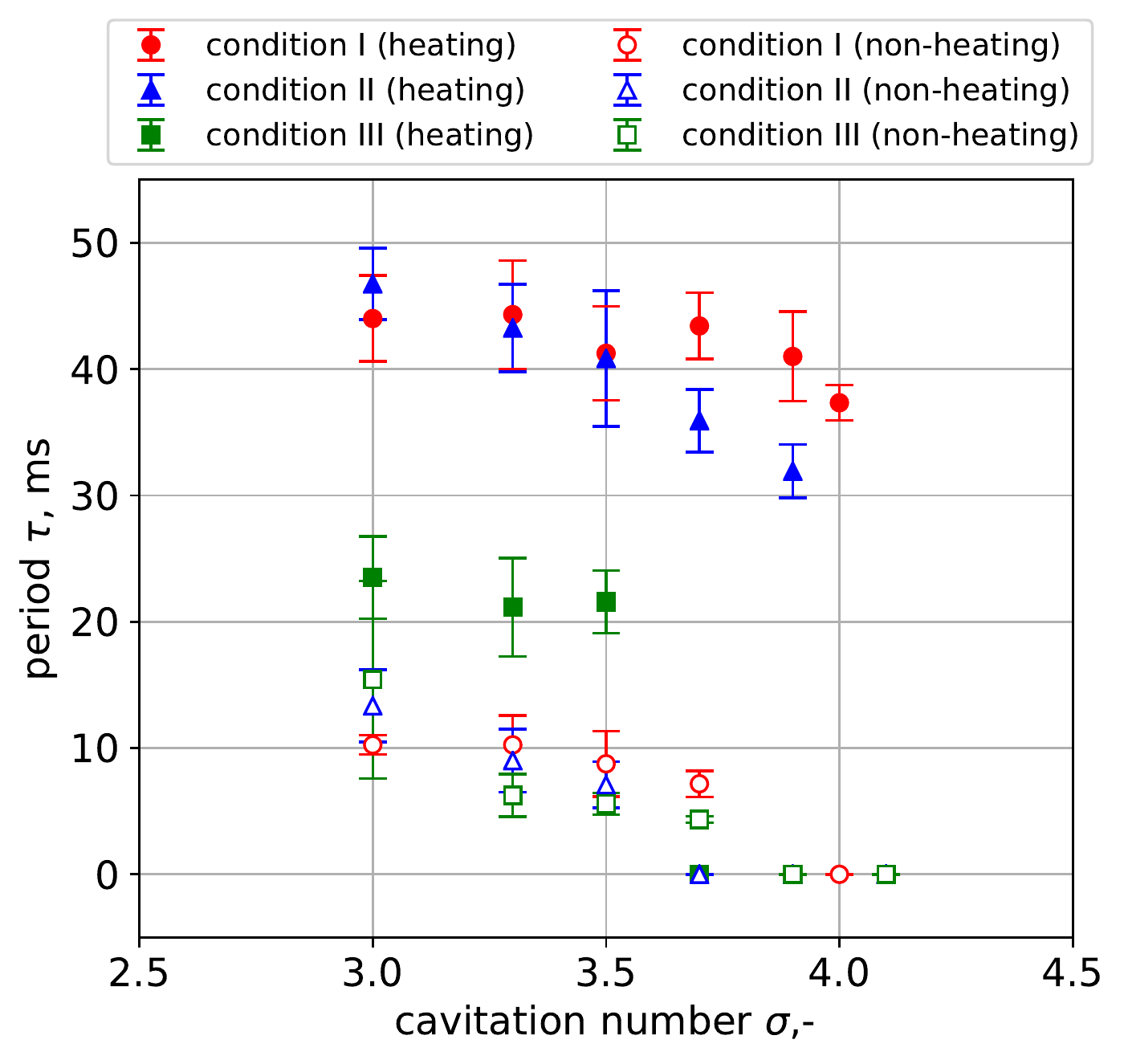}
\caption{Variation in periodicity with cavitation number under heating and 
non-heating conditions}
\label{fig_period}
\end{figure}

Figure \ref{fig_period} shows the variation in period with the cavitation number in the heating and non-heating cases.
A period of zero represents that the phenomenon was temporarily stable.
The period in the non-heating case increased as the cavitation number decreased, and the value of the period was approximately 10 ms. 
In contrast, in conditions I and II, the period was approximately 40 ms and a clear difference was observed between the heating and non-heating cases.
In addition, in Condition III, the period in the heating case increased when the cavitation number was lower than 3.5, although there was no difference between the cavity length for the heating and non-heating cases in this condition.

\section{Discussion}
\subsection*{Influence of heating on cavity aspect}
\begin{figure}[htbp] 
\centering
	\includegraphics[width=6.0in]{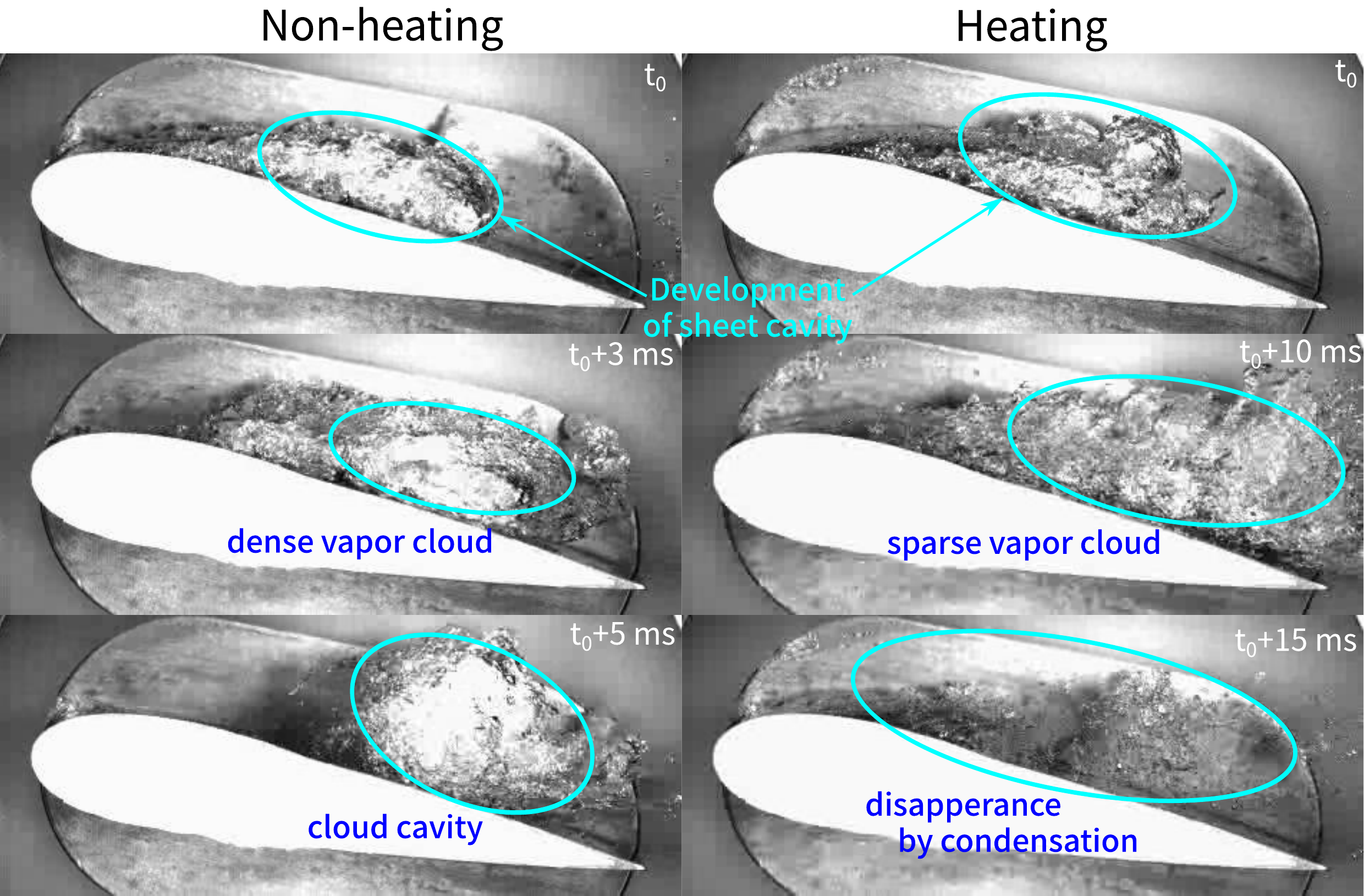}
	\caption{Typical difference in appearance of cavity (Condition I, 
$\sigma$ = 3.0)}
\label{fig_aspect}
\end{figure}

Figure \ref{fig_aspect} shows the typical appearance of the cavity in the heating and non-heating cases.
It should be noted that the pressure distributions in the heating and non-heating cases were almost similar when the cavity disappeared.
Therefore, the difference between the cavity appearance in the two cases was caused by the wall temperature; the temperature difference was 22.7 K.

As shown in the left column of Fig. \ref{fig_aspect}, typical sheet/cloud cavitation was observed in the non-heating case. 
This phenomenon consisted of cavity initiation, development of the sheet cavity, generation of the re-entrant jet from the trailing edge to the leading edge, breakoff of the sheet cavity, shedding of the cloud cavity, and collapse of the vapor cloud by pressure recovery at the wake of the hydrofoil.

As shown in the right column of Fig. \ref{fig_aspect}, the appearance of the cavity was completely different in the heating case.
First, the lifetime of the vapor cloud on the hydrofoil was considerably longer than that in the non-heating case, as also shown in Fig.  \ref{fig_period}.
This was caused by the difference in the vapor amount generated by evaporation.
As shown in Fig. \ref{fig_L_heat}, heating increased the cavity length. 

Additionally, as a general trend of the difference, the cavity in the heating case resembled a sparse vapor cloud while that in the non-heating case appeared dense.
Therefore, the size of bubbles in the vapor cloud is expected to be smaller in the heating case.
Furthermore, because the evaporative mass on the hydrofoil surface is increased by the heated hydrofoil, the cavity is longer in the heating case. 
However, the elongated part of the cavity in the heating case cannot stably exist from the viewpoint of the pressure field around the hydrofoil because it is generated forcibly by heating.
Moreover, the vapor phase near the hydrofoil is expected to grow larger and gain enthalpy by heat supply from the heated hydrofoil, but the vapor at the mainstream side will likely lose enthalpy owing to mainstream flow. 
Therefore, the elongated part will be collapsed easily by a relatively high-pressure field, thereby forming a sparse cloud cavity with smaller bubbles.

\subsection*{Comparison of heating effect at same cavity length}
\begin{figure}[htbp] 
\centering
	\includegraphics[width=6.0in]{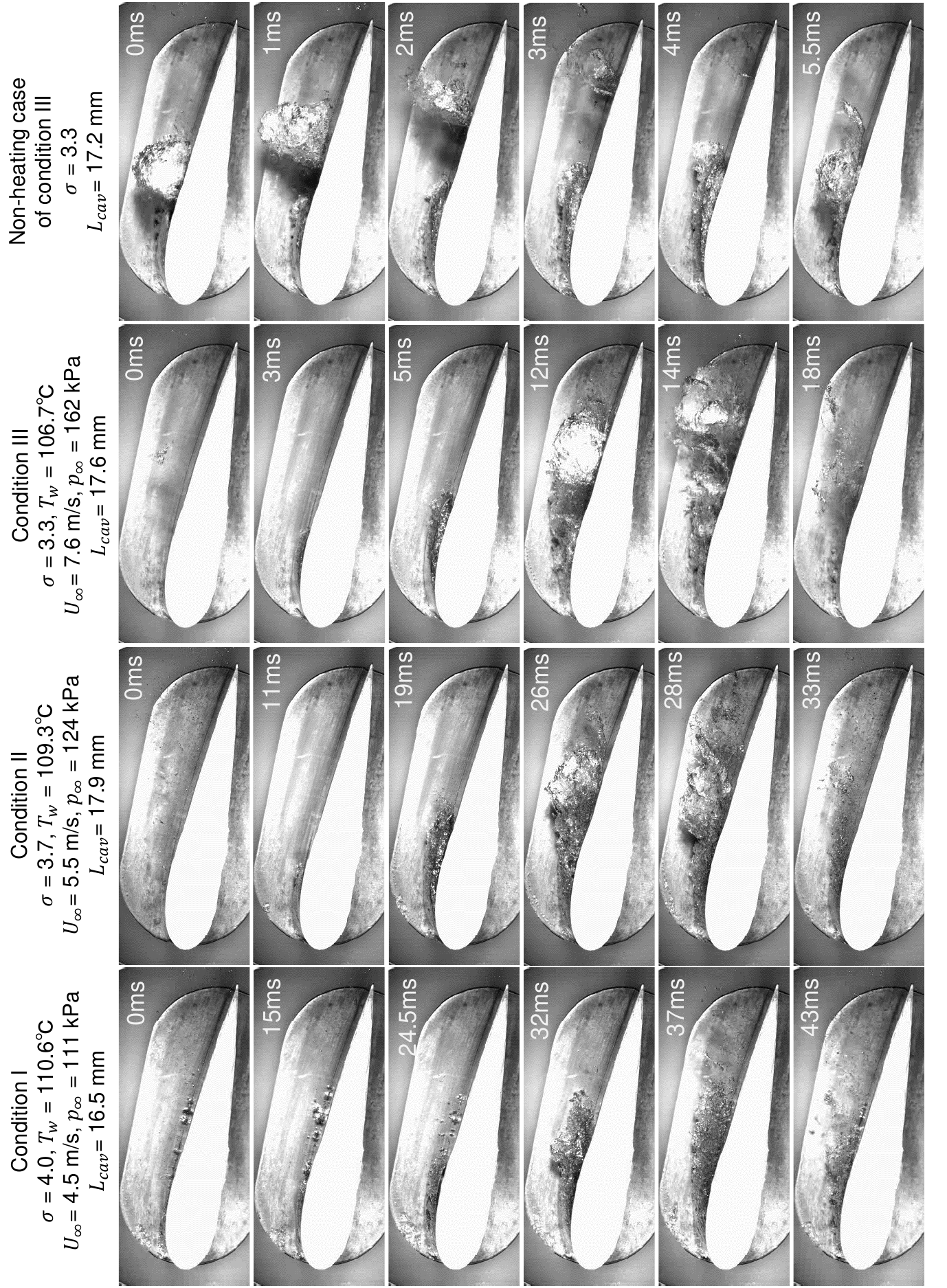}
\caption{Cavity aspects in the conditions with similar cavity lengths}
\label{fig_new_sigma_flow_pattern}
\end{figure}

Figure \ref{fig_new_sigma_flow_pattern} shows the comparison of the cavity aspects at similar cavity lengths and cavitation numbers, namely, similar vapor volume.
The images in the third row show the times at which the cavity length was measured.
The maximum length of sheet cavities was almost the same.  Although a similar vapor volume was observed, the origin of vapor generation in each condition was different.
At a low $\sigma$, the vapor was generated by local depressurization. In contrast, at a high $\sigma$, the vapor was partially generated by heating from the hydrofoil.
As shown in Figs. \ref{fig_new_sigma_flow_pattern} and \ref{fig_period}, the cavitation phenomena in the heating case had a longer periodicity and a more stable pattern than those in the non-heating case.
The longer periodicity was caused by the reduction in partial cavity instability and the achievement of a quasi-steady balance between continuous vapor supply from the heated hydrofoil and condensation in the upper region of the cavity.
As shown in the images for the non-heating case in Fig.  \ref{fig_new_sigma_flow_pattern}, the cloud cavity was released because of the re-entrant jet generated by the adverse pressure gradient at the end of the cavity\citep{CALLENAERE2001}.
On the contrary, because of the heating effect, the lifetime of the vapor bubble increased compared to the non-heating case, and the bubble could travel downstream.
These different behaviors of the cavity resulted in different flow patterns and periodicities.

As shown in Fig. \ref{fig_new_sigma_flow_pattern}, even though there were no differences between the cavity length and cavity pattern in Condition III, the periodicity in the heating case was longer than that in the non-heating case.
Additionally, as shown in Fig. \ref{fig_period}, the periodicity in the heating case in Condition III varied at a certain $\sigma$. This abrupt variation indicated that the heating effect explicitly occurred by overcoming a potential threshold between isothermal cavitation and cavitation with a heating effect.
In summary, the experimental results obtained in this study imply that the heating effect reduces the unsteadiness of cavitation.

\subsection*{Modification of cavitation number including wall heat transfer}

\begin{figure}[htbp] 
\centering
	\includegraphics[width=2.5in]{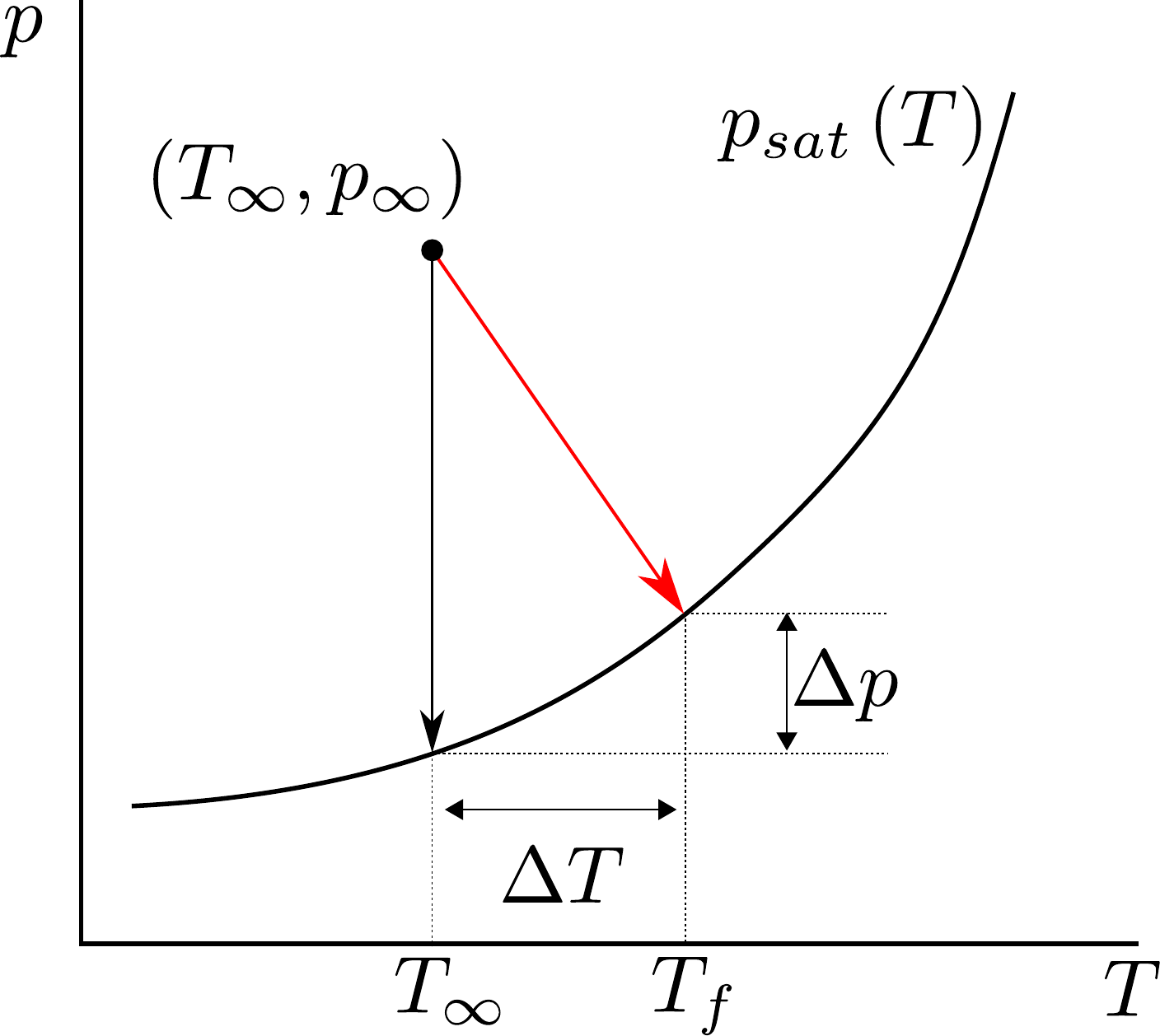}
\caption{Trajectory of phase change process under heating condition}
\label{fig_sat_curve}
\end{figure}

The modification of the cavitation number was examined.
We considered the effect of the hydrofoil temperature and mainstream velocity on the driving force for cavity development to derive the correction term for the cavitation number.
As shown in Fig. \ref{fig_L_heat}, the cavity length depended on the heat transfer from the hydrofoil. 
We assumed that the saturation pressure of the fluid close to the heated wall was locally modified, and this promoted evaporation. 
Figure \ref{fig_sat_curve} shows the phase change process on the saturation diagram.
Unlike the isothermal case, the intersection on the saturation curve shifted to a high temperature.
The cavitation number can be considered as the projection length between the inlet condition and this intersection on the pressure axis.
Therefore, the corrected cavitation number is defined by Eq.  (\ref{eq_sigma_star}),

\begin{align}
	\sigma^{*} &= \frac{p_\infty-p_{sat} \left( T_\infty \right)-\Delta 
p}{\frac{1}{2}\rho_L \left( T_\infty \right) U_\infty^2} \notag \\
	&\equiv \sigma - \sigma_T,
	\label{eq_sigma_star}
\end{align}

\noindent
where $\Delta p$ denotes the saturation pressure variation due to the heated hydrofoil.
$\Delta p$ is approximately expressed as the gradient of the saturation curve, as given by Eq. (\ref{eq_satP}),

\begin{equation}
	\Delta p \thicksim \frac{d p_{sat}}{dT} \Delta T.
	\label{eq_satP}
\end{equation}

\noindent
Equation (\ref{eq_satP}) shows the relationship between the corrected value of the saturation pressure and fluid temperature. 
Hence, the term used to correct the cavitation number, $\sigma_T$, which is referred to as the thermal correction term in this study, can be expressed as follows:

\begin{equation}
	\sigma_T \thicksim \frac{1}{\rho_L \left( T_\infty \right) 
U_\infty^2} \frac{d p_{sat}}{dT} \Delta T.
	\label{eq_sigmaT}
\end{equation}

\begin{figure}[htbp]
\centering
	\includegraphics[width=4.34in]{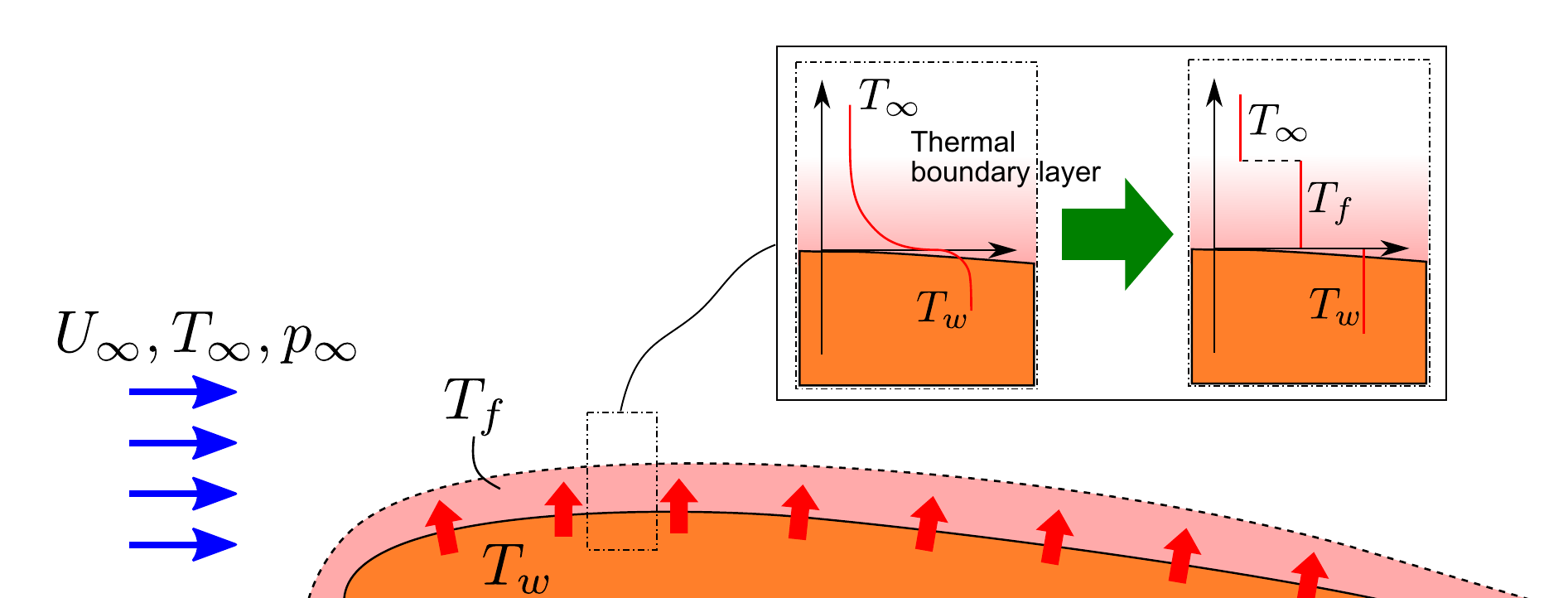}
\caption{Simplification of the treatment of thermal boundary layer around 
heated hydrofoil}
\label{fig_assumption}
\end{figure}

Next, we derived the relationship between the fluid temperature and hydrofoil temperature.
Figure \ref{fig_assumption} shows the simplification of the treatment of the thermal boundary layer around a heated hydrofoil. 
A thin thermal boundary layer is formed around the hydrofoil because of turbulent convective heat transfer. 
Even though the fluid temperature varies in the thermal boundary layer, we assume a spatially-temporally average temperature of the heated fluid layer, $T_f$, as shown in Fig. \ref{fig_assumption}.
This temperature is expressed by Eq. (\ref{eq_fluid_temp}),

\begin{equation}
	T_f = T_\infty + \frac{\dot{Q}}{\dot{m} c_{p} \left( T_\infty \right)},
	\label{eq_fluid_temp}
\end{equation}

where $\dot{Q}$ and $\dot{m}$ denote the total heat transfer rate and mass flow rate, respectively.
Here, the surface area of the suction side of the hydrofoil is assumed as the product of chord length $C$ and the span width $W$, total heat transfer $\dot{Q}$ is expressed as,

\begin{align}
	\dot{Q} &= h \left( T_w -T_\infty \right) C W \notag \\
	&= \nNu \: k_L \left( T_\infty \right) \left( T_w -T_\infty \right) W,
	\label{eq_Nu}
\end{align}

\noindent
where $h$ and $\nNu$ denote the heat transfer coefficient and Nusselt number, respectively. The characteristic length of convective heat transfer was assumed as the chord length $C$.
As shown in Fig. \ref{fig_assumption}, limited flow on the heated surface is affected by heating in this model.
Here, the effective height of the affected fluid layer is defined as $\delta_T^*$, then the mass flow rate is assumed as,
\begin{equation}
	\dot{m} = \rho_L \left( T_\infty \right) U_\infty W \delta_T^*.
	\label{eq_mfr}
\end{equation}

\noindent
Moreover, as the velocity in this layer has a large gradient, 
it is difficult to determine its thickness beforehand. Therefore, this parameter was treated as the experimental parameter.
Substituting Eqs. (\ref{eq_Nu}) and (\ref{eq_mfr}) into Eq. (\ref{eq_fluid_temp}), $\varDelta T$ shown in Fig. \ref{fig_sat_curve} is derived as follows,

\begin{equation}
	\varDelta T = T_f - T_\infty 
	            = \eta \frac{\nNu k_L \left( T_w - T_{\infty} \right)}
	                        {\rho_L c_p U_\infty C}  
	\label{eq_dT}
\end{equation}
\noindent
Here, it is assumed that the thickness of the affected fluid layer is represented as $\delta^*_T = C/\eta$ and $\eta$ is the experimental parameter.
Finally, substituting Eqs. (\ref{eq_dT}) into Eq. (\ref{eq_sigmaT}), the thermal correction term for the cavitation number is obtained as

\begin{equation}
	\sigma_T = \eta \frac{\nNu \: \alpha_L}{\rho_L U_\infty^3 C} 
\frac{d p_{sat}}{dT} \left( T_w - T_\infty \right),
	\label{eq_finalexp}
\end{equation}

\noindent
where coefficient $\eta$ is a fitting parameter.
Generally, the Nusselt number for turbulent flow is proportional to $\nRe^{0.8}$. 
Therefore, $\sigma_T$ is proportional to $U_\infty^{-2.2}$ and $T_w$.
The following correlation of turbulent convective heat transfer on a flat plate~\citep{Incropera2013} is used to calculate the actual value of the thermal correction term:

\begin{equation}
	\nNu = 0.037 \nPr^{\frac{1}{3}} \nRe^{\frac{4}{5}}.
	\label{eq_PrRe}
\end{equation}

\noindent
Equation (\ref{eq_finalexp}) can be solved using the inlet condition ($T_\infty$, $U_\infty$), measured value ($T_w$), and thermophysical properties evaluated using $T_\infty$ and $p_\infty$.

\begin{figure}[htbp]
\centering
	\includegraphics[width=3.34in]{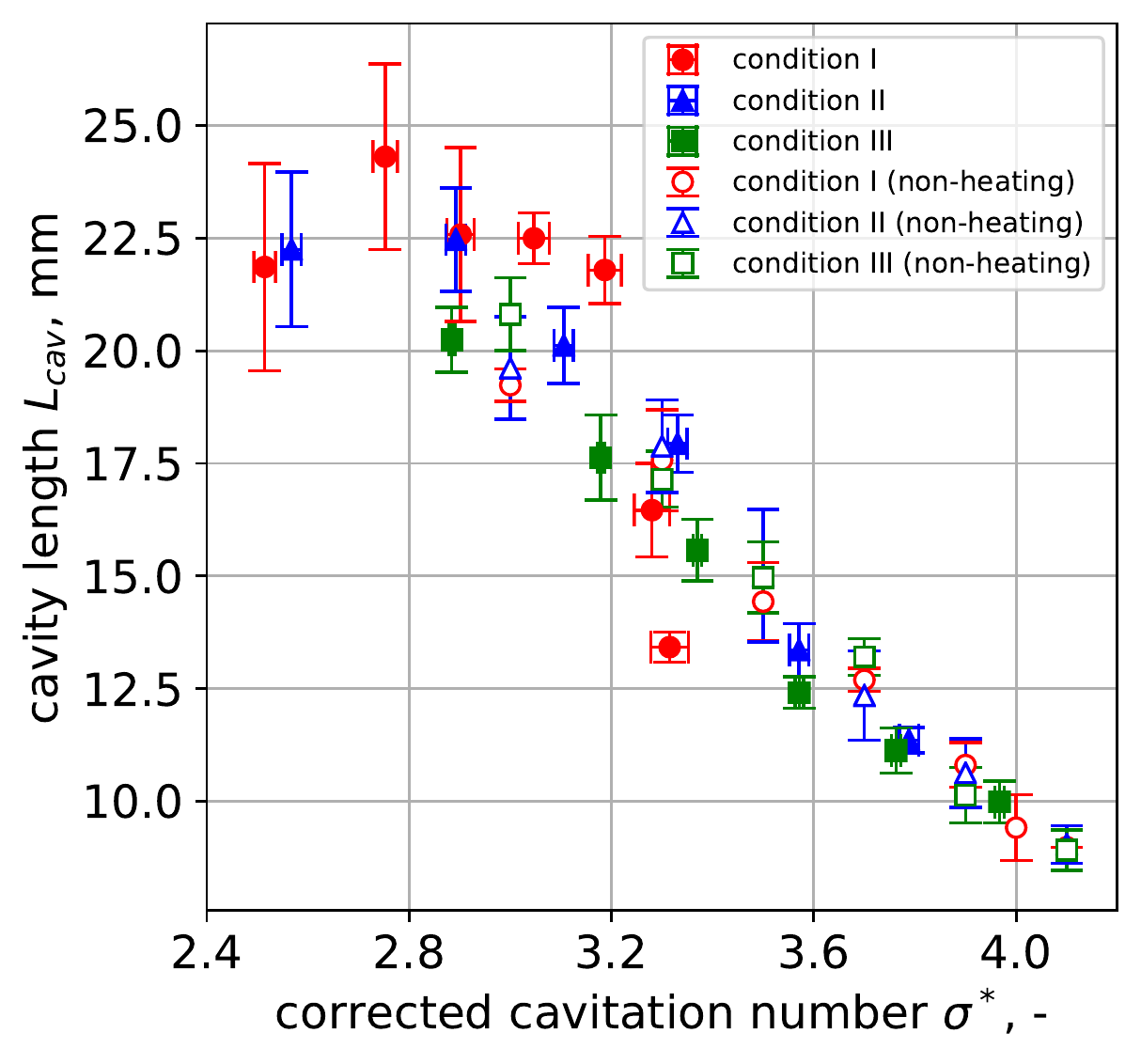}
\caption{Relationship between corrected cavitation number and cavity length 
in non-heating and heating conditions with $\eta$ = 120. Hollow and solid 
symbols denote non-heating and heating conditions, respectively}
\label{fig_new_sigma}
\end{figure}

Figure \ref{fig_new_sigma} shows the relationship between the cavity length and corrected cavitation number.
$\eta$ was set as 120.
The experimental results for all conditions were in agreement with each other.
In addition, as explained in Fig. \ref{fig_new_sigma_flow_pattern}, the flow pattern and periodicity were different even for the same cavity length.

\begin{figure}[htbp]
\centering
	\includegraphics[width=3.34in]{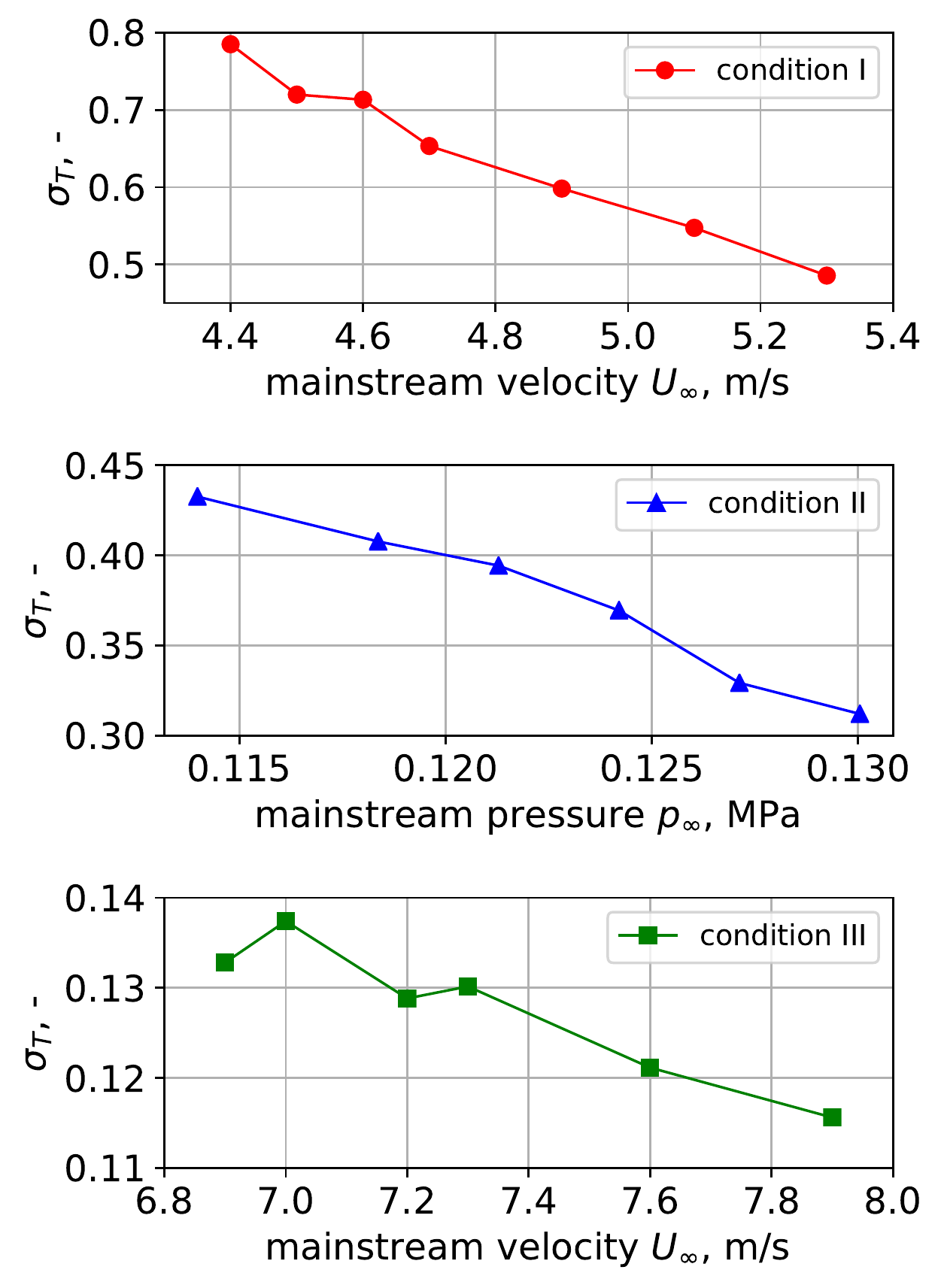}
\caption{Values of $\sigma_T$ in each condition, calculated using $\eta$ = 
120}
\label{fig_All_sigT}
\end{figure}

Figure \ref{fig_All_sigT} shows the values of $\sigma_T$ in each condition, with $\eta$ set as 120.
The thermal correction term accurately described the trend of the thermal effect in the cavitating flow around the heated hydrofoil.
In all conditions, the thermal correction term decreased as the mainstream velocity and pressure increased.
The thermal correction term in Condition I was the largest because of the high hydrofoil temperature and low mainstream velocity.
Furthermore, the thermal correction term for Condition III indicated that the heating effect in this condition was smaller than that in other conditions.

\section{Conclusions}
The influence of heat transfer from a heated object on cavitating flow was evaluated by performing a water tunnel experiment using a heated NACA0015 hydrofoil. The following results were obtained:
\begin{enumerate}
	\item The heating effect on cavitating flow was confirmed. The cavity length, cavity aspect, and periodicity were affected by the heat transfer from the hydrofoil.
	\item The heating effect became stronger at a lower velocity. This is in contrast to the trend observed in isothermal cavitation, where cavities grow as velocity increases.
	\item The periodicity of cavitating flow increased because of the heating effect.
	The periodicity in the heating case was longer than that in the non-heating case, even though there was no difference between the cavity length and cavity pattern. 
	\item A corrected cavitation number was derived by considering the influence of the heating effect on the driving force for cavity development. 
        The corrected cavitation number was evaluated using the experimental data for the cavity length. Consequently, the cavity lengths for the non-heating and heating cases were expressed by a unified parameter. 
\end{enumerate}

\section*{Funding}
This research received no specific grant from any funding agency in the public, commercial, or not-for-profit sectors.

\bibliography{draft}

\end{document}